\documentclass[10pt, oneside]{article}   	
\usepackage{geometry} 
\usepackage{graphicx}               
\usepackage{graphicx}				
\usepackage{amsmath}
\usepackage{cite}
\newtheorem{definition}{Definition}[section]
	
\newtheorem{theorem}{Theorem}[section]	
	
\newtheorem{proof}{Proof}[section]							
\usepackage{amssymb}
\usepackage{indentfirst}
\title{SlideVaR: a risk measure with variable risk attitudes}
\author{Hu Wentao}

\date{April 23, 2019}
\begin{document}
\maketitle
\begin{abstract}
	To find a trade-off between profitability and prudence, financial practitioners need to choose appropriate risk measures. Two key points are: Firstly, investors' risk attitudes under uncertainty conditions should be an important reference for risk measures. Secondly, risk attitudes are not absolute. For different market performance, investors have different risk attitudes. We proposed a new risk measure named SlideVaR which fully reflects the different subjective attitudes of investors and sufficiently reflects the impact of market changes on investor attitudes. We proposed the concept of risk-tail region and risk-tail sub-additivity and proved that SlideVaR satisfies several important mathematical properties. Moreover, SlideVaR has a simple and intuitive form of expression for practical application. Several simulate and empirical computations show that SlideVaR has obvious advantages in markets where the state changes frequently.\par
~	\par
Keywords: SlideVaR; risk-tail region; risk-tail sub-additivity; risk attitudes; VaR; CVaR
\end{abstract}

\section{Introduction}
	Risk management is a problem that investors face every day. Investors need to find a trade-off between profitability and prudence. In the process of achieving this purpose, choosing appropriate risk measures plays an important role. One popular risk measure is VaR. Let $X$ be the loss,
\begin{eqnarray}
VaR_{\alpha}(X) = inf\{x\in \mathbb{R}:P[X\leqslant x]> \alpha\} = F^{-1}_{X}(\alpha).
\end{eqnarray} 	
Simplicity is an important advantage. First of all, VaR is easy to understand. According to Duffie and Pan\cite{duffie1997}, VaR can be defined as For a given time horizon $T$ and a confidence level $\alpha$, VaR is the loss in market value that can only be exceeded with a probability of at most $1-\alpha$. VaR is simply the $\alpha$ percentile of the loss distribution. In addition, VaR is easy to calculate and get back-test\cite{Kupiec1995,Christoffersen1998,Hull2012}. However, VaR can only give a threshold under a certain probability, but cannot describe the losses which exceed the threshold. Consequently, catastrophic losses may be seriously underestimated especially when losses distribution tend to exhibit fat-tail. Moreover, VaR does not satisfy sub-additivity. Therefore VaR may overestimate the overall risk by ignoring the diversity among different parts of a portfolio\cite{Perignon2008,Perignon2010}.\par 
	CVaR was proposed by Rockafellar in 1997 as an improved method.
\begin{eqnarray}
CVaR_{\alpha}(X) = \frac{1}{1-\alpha}\int_{\alpha}^{1}F^{-1}_{X}(p) dp = \mathbb{E}[X|X\geqslant VaR_{\alpha}(X)].
\end{eqnarray}	 
CVaR is the conditional expectation that represents the mean of extreme losses that exceed VaR. CVaR meets more attractive mathematical properties. Coherent risk measure was proposed by Artzner et al.\cite{Artzner1999}: A risk measure satisfying the translation invariance, sub-additivity, positive homogeneity, and monotonicity is called coherent. They proved that CVaR satisfies sub-additivity and is coherent. However, CVaR still ignores the human's subjective risk attitude. From the perspective of risk perception, whether the financial risk will cause damage and the scale of damage are related to the subjective perception of human beings and the objective existence of risk. Therefore, investors' risk attitudes under uncertainty conditions should be an important reference for risk measures. \par
	Wang\cite{Wang1995,Wang1997} first proposed distortion risk measure in 1997. Given a distortion function $g$, the distortion risk measure $\rho_{g}(X)$ is:
\begin{eqnarray}
\rho_{g}(X) = \int_{-\infty}^{0}[g(P[X > x])-1]dx + \int_{0}^{+\infty}g(P[X > x])dx.
\end{eqnarray}
(Ref ...) denoted that distortion risk measure is coherent if and only if the distortion function is concave. Spectral risk measure was proposed by Acerbi\cite{acerbi2001}. Given a risk aversion function $\phi$, We call the function
\begin{eqnarray}
M_{\phi}(X):=\int_0^1F^{-1}_X(p)\phi(p)dp, X\in\mathcal{X}
\end{eqnarray}
the spectral risk measure. Spectral risk measure is obtained by weighting different quantiles by $\phi$. Acerbi\cite{acerbi2001} proved that spectral risk measure is coherent. Distortion function $g$ and risk aversion function $\phi$ appears as the instrument by which an investor can express his subjective attitude. GlueVaR was proposed by Belles-Sampera\cite{Belles-Sampera2014a} to devise a risk measure that lies somewhere between VaR and CVaR. It is a special distortion risk measure and can be expressed as a linear combination of VaR and CVaR with two confidence levels:
\begin{eqnarray}
GlueVaR_{\alpha,\beta}(X) =\omega_1\cdot VaR_{\alpha}(X)+ \omega_2\cdot CVaR_{\alpha}(X)+\omega_3\cdot CVaR_{\beta}(X),\alpha<\beta.
\end{eqnarray}
However, because risk attitudes of these models are static, its can't reflect investors subjective attitudes sufficient enough. In the actual investment market, the risk attitudes of most investors are not absolute but will change with different situation. For example, when the market gets worse, investors tend to use a more conservative strategy. At this point they are more risk-averse. Conversely, if the current situation is good, some investors will adopt a more aggressive strategy, at which point they will tolerate higher risks. In other words, some investors would have an evaluation of the performance of the market and for different market performance, investors have different risk attitudes. Therefore, when measuring risk, the impact of the market environment on risk attitude is a dispensable factor.
\par 
	$\Lambda$VaR contains considerations for this factor. Frittelli\cite{Frittelli2014} proposed $\Lambda$VaR by replacing the constant confidence level $\alpha$ with a non-decreasing right continuous function $\Lambda(x)$:
\begin{eqnarray}
\Lambda VaR_{\alpha}(X) = inf\{x\in \mathbb{R}:P[X\leqslant x]> \Lambda(x)\}.
\end{eqnarray} 
$\Lambda$VaR is the minimum intersection of this function and the loss cumulative distribution function(CDF). In terms of risk attitudes, the tolerance level of $\Lambda$VaR will be modified as the loss distribution changes. In other words, the users of $\Lambda$VaR can only accept the larger loss occurred with lower probability. Although Frittelli\cite{Frittelli2014} considers the impact of the market environment on risk attitudes, the direct usage of the intersection of the loss CDF and $\Lambda(x)$ can't describe how does the market environment affect risk attitudes. Besides, whatever how the distribution changes, $\Lambda$VaR is still a quantile with certain confidence level. Therefore, $\Lambda$VaR and VaR may lead to the same distasteful outcomes in some extreme situations.\par
	Our motivation is to design a risk measure that fully reflects the different subjective attitudes of investors and sufficiently reflects the impact of market changes on investor attitudes. At the same time take into account some of the good mathematical properties. Finally, the new risk measure should have a relatively simple and intuitive form of expression for practical application. Therefore, we proposed a new risk measure named SlideVaR. Firstly, SlideVaR can be expressed as a combination of VaR and CVaR. With different weights, the risk attitude of SlideVar can slide between VaR and CVaR. Secondly, the weights of VaR and CVaR are determined by a tail-indicator $S(U_{\alpha}^{\phi}(X))$ which we defined to measure the thickness of the tail of the distribution. With the CDF of X changes, $S(U_{\alpha}^{\phi}(X))$ changes and then changes the risk attitude of SlideVaR. Thirdly, we proposed the concept of risk-tail region i.e. each element of this region is no less risky than those outside the region. Accordingly, we proposed risk-tail sub-additivity and prove that SlideVaR satisfies it. The subsequent work is structured as follows. In the second section, we introduce some relevant basic knowledge. In the third section, we define SlideVaR and analyze its mathematics properties. In the fourth section, some illustrations are provided. Finally, we discuss and conclude. \par
		
\section{Preliminaries}
\subsection{coherent risk measure}
Before we quantifying risk, supervisors should decide which kind of losses are unacceptable and next determine a minimum amount of risk reserve capital to make it's acceptable. This minimum reserve capital is called risk measure generally.  Risk measure can be defined as a mapping $\rho:\mathcal{X}\to \mathcal{R}$ where $\mathcal{X}$ is the collection of all possible losses. Coherent risk measure was proposed by Artzner et al.\cite{Artzner1999} which contains their risk attitudes towards assets management.\par
\begin{definition}[coherent]
A risk measure satisfying the four axioms of translation invariance, sub-additivity, positive homogeneity and monotonicity, is called coherent risk measure.

(a) Translation invariance: 
\begin{eqnarray}
\rho(X + a)=\rho(X)+a, \,\, \forall X \in \mathcal{X}, \,\, \forall a \in \mathbb{R}.
\end{eqnarray}

(b) Sub-additivity:
\begin{eqnarray}
\rho(X_1+X_2) \leqslant \rho(X_1)+\rho(X_2), \,\, \forall X_1, \,\, X_2 \in \mathcal{X}.
\end{eqnarray}

(c) Positive homogeneity:
\begin{eqnarray}
\rho(\lambda X)=\lambda \rho(X), \,\, \forall \lambda \geqslant 0, \,\, \forall X \in \mathcal{X}.
\end{eqnarray}

(d) Monotonicity:
\begin{eqnarray}
\mbox{if } X \leqslant Y, \mbox{then }  \rho(X) \leqslant \rho(Y), \,\, \forall X, \,\, Y \in \mathcal{X}.
\end{eqnarray}

\end{definition}\par
Artzner et al.\cite{Artzner1999} proved that VaR failed to satisfy sub-additivity and thus is not a coherent risk measure. In contrast, CVaR is coherent. \par
\subsection{Spectral risk measure}	
	Many classic risk measures can be obtained by weighting the quantiles of the loss distribution. For instance, $VaR_{\alpha}$ assigns $100\%$ weight to $\alpha\%$ quantile and $0\%$ weight to the others. $CVaR_{\alpha}$ is attained by assigning the same weight to the quantiles above $\alpha\%$ quantile yet $0\%$ to the remaining quantiles. Acerbi\cite{acerbi2001} promoted this idea and proposed spectral risk measure which allocates different weights for every quantile.\par 
	Firstly, Acerbi\cite{acerbi2001} defined the risk aversion function $\phi$ which is an element of the normed space $L^1([0,1])$ where every element is represented by a class of functions which differ at most on a subset of $[0,1]$ of zero measure. The norm in this space in given by	
\begin{eqnarray}
||\phi|| = \int_0^1|\phi(p)|dp.
\end{eqnarray}	
Evidently, monotonicity and positivity of an element of $L^1([0,1])$ cannot be defined pointwise as for functions. Hence,
\begin{definition}[monotonicity]
$\phi$ is decreasing if $\forall q \in (a,b)$ and $\forall \varepsilon >0$ such that $[q-\varepsilon,q+\varepsilon] \subset [a,b]$,
\begin{eqnarray}
\int_{q-\varepsilon}^{q}\phi(p)dp \geqslant \int_{q}^{q+\varepsilon}\phi(p)dp.
\end{eqnarray}
$\phi$ is increasing if $\forall q \in (a,b)$ and $\forall \varepsilon >0$ such that $[q-\varepsilon,q+\varepsilon] \subset [a,b]$,
\begin{eqnarray}
\int_{q-\varepsilon}^{q}\phi(p)dp \leqslant \int_{q}^{q+\varepsilon}\phi(p)dp.
\end{eqnarray}
\end{definition}	
\begin{definition}[positivity]
$\phi$ is positive if $\forall I \subset [a,b]$,
\begin{eqnarray}
\int_I\phi(p)dp \geqslant 0.
\end{eqnarray}
\end{definition}
The function $\phi$ is used to set weights for every quantile. In practice, $\phi$ appears as the instrument by which an investor can express his subjective attitude toward risk. \par
	Next, Acerbi\cite{acerbi2001} give the following,
\begin{definition}[spectral risk measure]
If a risk aversion function $\phi \in L^1([0,1])$ satisfies: (1) $\phi$ is positive, (2) $\phi$ is increasing, (3) $|| \phi ||=1$. Then we call the function
\begin{eqnarray}
M_{\phi}(X):=\int_0^1F^{-1}_X(p)\phi(p)dp, X\in\mathcal{X}
\end{eqnarray}
is the spectral risk measure generated by $\phi$.
\end{definition}
Acerbi\cite{acerbi2001} proved that spectral risk measure is coherent. The fact that $\phi$ is an increasing monotonic function is the key point of the proof of sub-additivity. This fact provides us with an intuitive insight of the concept of coherence:``a measure is coherent if it assigns bigger weights to worse case.''\par

\section{A new family of risk measures: SlideVaR}
\subsection{Basic structure of SlideVaR}
	For the purpose of convenience, we firstly give the basic structure of SlideVaR. Use the same notations as above section,
\begin{definition}
Given two confidence levels $\alpha$ and $\beta$, $\alpha \geqslant \beta$, for loss $X \in \mathcal{X}$,  $SlideVaR_{\alpha,\beta}^{\phi}(X)$ is defined by
\begin{eqnarray}
SlideVaR_{\alpha,\beta}^{\phi}(X) = S(U_{\beta}^{\phi}(X))\cdot CVaR_{\alpha}(X) + [1-S(U_{\beta}^{\phi}(X))]\cdot VaR_{\beta}(X)
\end{eqnarray}
where $U_{\beta}^{\phi}(X):\mathcal{X}\to \mathcal{R}$ is called tail thickness of $X$, and $S(x):[min(X),max(X)] \to [0,1]$ is a normalization function.
\end{definition}\par
By definition, $SlideVaR_{\alpha,\beta}^{\phi}$ is a combination of $CVaR_{\alpha}$ and $VaR_{\beta}$. The confidence levels $\alpha$ and $\beta$ have different meanings. The smaller one $\beta$ corresponds to the bad scenarios, and the larger one $\alpha$ corresponds to worst scenarios. The risk attitude of SlideVaR is controlled by $S(U_{\beta}^{\phi}(X))$. When $S(U_{\beta}^{\phi}(X))=0$, SlideVaR is the most aggressive. It represents the minimum loss of bad scenarios(i.e. $VaR_{\beta}$). When $S(U_{\beta}^{\phi}(X))=1$, SlideVaR is the most conservative. It represents the average loss of worst scenarios(i.e. $CVaR_{\alpha}$). When $S(U_{\beta}^{\phi}(X))\in(0,1)$, the risk attitude slides between the most aggressive and conservative one. Reviewing
\begin{eqnarray}
GlueVaR_{\alpha,\beta}(X) =\omega_1\cdot VaR_{\alpha}(X)+ \omega_2\cdot CVaR_{\alpha}(X)+\omega_3\cdot CVaR_{\beta}(X),
\end{eqnarray}
it's easy to find that SlideVaR and GlueVaR share the similar perspective. However, limited by the distortion risk measure framework, the coefficients of GlueVaR are constants. When the market is during a period with extreme high volatility, VaR tends to underestimate some catastrophic losses. At this point VaR part of the combination will impairs the ability of GlueVaR to cover the risk. Conversely, because CVaR is more conservative than VaR, CVaR part will cause GlueVaR over inflated in the low or mild volatility period. Considering $\Lambda$VaR,
\begin{eqnarray}
\Lambda VaR_{\alpha}(X) = inf\{x\in \mathbb{R}:P[X\leqslant x]> \Lambda(x)\}.
\end{eqnarray}
Because $\Lambda$VaR is the minimum VaRs with a set of confidence levels, whatever how the loss distribution changes, $\Lambda$VaR is still a quantile with certain confidence level. This means that $\Lambda$VaR can only be adjusted from the confidence level aspect. Therefore, $\Lambda$VaR and VaR may lead to the same distasteful outcomes in some extreme situations. On the contrary, ability of SlideVaR to cover extreme risks will increase as the proportion of $CVaR_{\alpha}(X)$ increases.\par
	It can be seen that $U_{\alpha}^{\phi}(X)$ is the key to the variable risk attitudes of SlideVaR. It also directly affects the performance of SlideVaR in different market conditions.\par
\subsection{Tail thickness: $U_{\alpha}^{\phi}(X)$}
	In this section, we define an indicator $U_{\beta}^{\phi}$ to depict features of loss $\beta$-tail distribution. It is not only the basis for the adjustment of risk attitude, but also fully reflects the subjective attitude of practitioners.
\begin{definition}
Given the confidence level $\beta$, the indicator $U_{\beta}^{\phi}$ is defined by:
\begin{eqnarray}
U_{\beta}^{\phi}(X):=\int_0^1F^{-1}_X(p)\phi_{\beta}(p)dp.
\end{eqnarray}
Function $\phi_{\beta} \in L^1([0,1])$ is
\begin{eqnarray}
\phi_{\beta}(p) = \left\{
\begin{aligned}
& 0  					&\,\, \text{if }\,\, & p < \beta\\
& \phi(p) 				&\,\, \text{if }\,\, & p \geqslant \beta
\end{aligned} \right.
\end{eqnarray}
where $\phi \in L^1([F_{X}^{-1}(\beta),1])$ satisfies: (1) $\phi$ is positive, (2) $\phi$ is non-decreasing, (3) $|| \phi ||=1$. 
\end{definition}
The non-decreasing function $\phi_{\beta}$ indicates that, for the quantiles which below the $\beta\%$ quantile, we give them 0 weight. On the contrary, we give greater weights to larger quantiles which above the $\beta\%$ quantile. As the result, regarding two losses $X_1,X_2 \in \mathcal{X}$ satisfying 
\begin{eqnarray}
\left\{
\begin{aligned}
& F_{X_1}^{-1}(p) \ne F_{X_2}^{-1}(p)    	&\,\, \text{if }\,\, & p < \beta,\\
& F_{X_1}^{-1}(p) = F_{X_2}^{-1}(p)  	&\,\, \text{if }\,\, & p \geqslant \beta,
\end{aligned} \right.
\end{eqnarray}
we have 
\begin{eqnarray}
U_{\beta}^{\phi}(X_1) = U_{\beta}^{\phi}(X_2).
\end{eqnarray}
Therefore, $U_{\beta}^{\phi}(X)$ is designed specifically to describe the features of $\beta$-tail distribution. We call the tail-indicator $U_{\beta}^{\phi}(X)$ the tail thickness of $X$.\par
In point of fact, $U_{\beta}^{\phi}$ is a spectral risk measure. According to monotonicity, we have
\begin{eqnarray}
U_{\beta}^{\phi}(X) \in [min(X),max(X)].
\end{eqnarray}
Considering the structure of SlideVaR, $U_{\beta}^{\phi}(X)$ is used to determine the weights of the elements in the combination. Hence, we need define a normalization function to convert $U_{\beta}^{\phi}(X)$ to a value between 0 to 1.
\begin{definition}
We call the non-decreasing function 
\begin{eqnarray}
S(x): [min(X),max(X)] \to [0,1]
\end{eqnarray}
the normalization function $S(x)$.
\end{definition}
Combining $U_{\beta}^{\phi}(X)$ and $S(x)$, we use $S(U_{\beta}^{\phi}(X))$ as the weight of $CVaR_{\alpha}(X)$ in $SlideVaR_{\alpha,\beta}^{\phi}(X)$ expression. Further, because $S(x) \in [0,1]$, 
\begin{eqnarray}\label{mon_2}
VaR_{\beta}(X) \leqslant SlideVaR_{\alpha,\beta}^{\phi}(X) \leqslant CVaR_{\alpha}(X).
\end{eqnarray}
This is the reason why $U_{\beta}^{\phi}(X)$ is designed specifically for $\beta$-tail distribution: regardless of the shape of $\beta$ left side distribution, the corresponded risk is covered by all possible minimums of SlideVaR.\par
	$S(U_{\beta}^{\phi}(X))$ satisfies following monotonicity properties:
\begin{theorem}\label{mon_U}(Mon1): $\forall X_1 \geqslant X_2 \in \mathcal{X}$, 
\begin{eqnarray}
S(U_{\beta}^{\phi}(X_1)) \geqslant S(U_{\beta}^{\phi}(X_2)).
\end{eqnarray}\par
(Mon2): If $a \in \mathcal{R}$ and $a \geqslant 0$, then
\begin{eqnarray}
S(U_{\beta}^{\phi}(X+a)) \geqslant S(U_{\beta}^{\phi}(X)).
\end{eqnarray}
If $a \leqslant 0$, then
\begin{eqnarray}
S(U_{\beta}^{\phi}(X+a)) \leqslant  S(U_{\beta}^{\phi}(X)).
\end{eqnarray}\par
(Mon3): If $\lambda \in \mathcal{R}$ and $\lambda \geqslant 1$, then
\begin{eqnarray}
S(U_{\beta}^{\phi}(\lambda X)) \geqslant \lambda S(U_{\beta}^{\phi}(X)).
\end{eqnarray}
If $0 \leqslant \lambda \leqslant 1$, then
\begin{eqnarray}
S(U_{\beta}^{\phi}(\lambda X)) \leqslant \lambda S(U_{\beta}^{\phi}(X)).
\end{eqnarray}
\end{theorem}
The above theorem is a simple corollary of monotonicity of $U_{\beta}^{\phi}(X)$ and $S(x)$.\par
	Because CVaR is more conservative than VaR and $\alpha>\beta$, these monotonicity properties reflect that as the scale of loss increases, the conservation degree of SlideVaR increases. For different $X$, the function $\phi$ is fixed. By defining the function $\phi$, investors can express their views on the tail of the distribution in a position that is independent of market performance. Therefore, the tail thickness $U_{\beta}^{\phi}(X)$ can be regarded as the investors' evaluation of the market. The function $S$ reflects how investors' market evaluation affects investors' risk attitudes. For example, during an economic crisis, we may prefer to use a very conservative strategy. Because panic may make us very sensitive to any turmoil. Conversely, when the economy is booming, to earn more profits, we may tend to use more aggressive strategies with higher tolerance towards losses.\par 	
\subsection{Investors' subjective attitude}
	SlideVaR considers the subjective attitudes of practitioners sufficiently. Because $U_{\beta}^{\phi}(X)$ is designed to describe the feature of distribution by a single number, it can be regarded as an investor's evaluation of market situation. $\phi_{\beta}(p)$ can be regarded as a ``rule'' which is used to determine the proportion of a certain quantile in this ``evaluation''. In fact, $\phi_{\beta}(p)$ is both objective and subjective. On the one hand, $\phi_{\beta}(p)$ is a function of $p\in [0,1]$, which is independent of the distribution of loss $X$. Namely, Once the $\phi_{\beta}(p)$ is determined, the ``rules of evaluation'' will not be affected by market changes. On the other hand, the ``rules'' are completely determined by the subjective attitude of investors. For instance, \par
	(a) In $[\beta,1]$, $\phi_{\beta}(p)$ is convex. This means that investors' assessment of the market is more sensitive to huge losses. Here are a few special examples: Exponential function
\begin{eqnarray}
\phi_{\beta}(p) = \left\{
\begin{aligned}
& 0  					&\,\, \text{if }\,\, & p < \beta\\
& \dfrac{e^{\frac{p-1}{\gamma}}}{\gamma(1- e^{\frac{\beta-1}{\gamma}})}	&\,\, \text{if }\,\, & p \geqslant \beta
\end{aligned} \right.
\end{eqnarray}	
where $\gamma \in [0,1]$ is a constant. Power function	
\begin{eqnarray}
\phi_{\beta}(p) = \left\{
\begin{aligned}
& 0  					&\,\, \text{if }\,\, & p < \beta\\
& \dfrac{(1-\gamma)(1-p)^{-\gamma}}{(1-\beta)^{1-\gamma}}	&\,\, \text{if }\,\, & p \geqslant \beta
\end{aligned} \right.
\end{eqnarray}	
where $\gamma \in [0,1]$ is a constant. Hans\cite{Hans} established relationship between utility function and spectral risk measures. The exponential spectral function and power spectral function can be derived from the exponential utility function and power function respectively. The above two functions $\phi_{\beta}(p)$ are obtained by simple transformation from these two special spectral functions.\par
	(b) In $[\beta,1]$, $\phi_{\beta}(p)$ is concave. This means that investors' assessment of the market is less sensitive to huge losses. For example: Power function	
\begin{eqnarray}
\phi_{\beta}(p) = \left\{
\begin{aligned}
& 0  					&\,\, \text{if }\,\, & p < \beta\\
& \dfrac{(1+\gamma)p^{\gamma}}{1-\beta^{1-\gamma}}	&\,\, \text{if }\,\, & p \geqslant \beta
\end{aligned} \right.
\end{eqnarray}	
where $\gamma \in [0,1]$ is a constant.\par
	(c) In $[\beta,1]$, $\phi_{\beta}(p)$ is linear. This means that the sensitivity of investors' assessment will not change. The spectral function of CVaR is a special case:
\begin{eqnarray}
\phi_{\beta}(p) = \left\{
\begin{aligned}
& 0  					&\,\, \text{if }\,\, & p < \beta\\
& \frac{1}{1-\beta}			 &\,\, \text{if }\,\, & p \geqslant \beta
\end{aligned} \right. .
\end{eqnarray}\par
	(d) In $[\beta,1]$, $\phi_{\beta}(p)$ is a general function. For example, $\phi_{\beta}(p)$ is a step function 
\begin{eqnarray}
\phi_{\beta}(p) = \left\{
\begin{aligned}
& 0  					&\,\, \text{if }\,\, & p < \beta\\
& \frac{\omega_1 p}{1-\beta}			 &\,\, \text{if }\,\, & \beta \leqslant p < \beta_1\\
& \frac{\omega_1 p}{1-\beta}+\frac{\omega_2 p}{1-\beta_1}			 &\,\, \text{if }\,\, & \beta_1 \leqslant p < \beta_2\\
& \frac{\omega_1 p}{1-\beta}+\frac{\omega_2 p}{1-\beta_1}+\frac{\omega_3 p}{1-\beta_2}			 &\,\, \text{if }\,\, & p \geqslant \beta_2
\end{aligned} \right. 
\end{eqnarray}	
where $\beta < \beta_1 < \beta_2 < 1$, $\omega_1,\omega_2,\omega_3 \in[0,1]$ and $\omega_1+\omega_2,+\omega_3=1$. Using this $\phi_{\beta}(p)$, $U_{\beta}^{\phi}(X)$ can be express as
\begin{eqnarray}
U_\beta^{\phi}(X) = \omega_1\cdot CVaR_{\beta}(X)+\omega_2\cdot CVaR_{\beta_1}(X)+\omega_3\cdot CVaR_{\beta_2}(X).
\end{eqnarray}\par
	The function $S$ reflects how investors' market evaluation affects investors' risk attitudes. If $S$ is convex, with tail thickness $U_{\beta}^{\phi}(X)$ increases, the degree of conservation of SlideVaR accelerates. If $S$ is concave, the degree of conservation of SlideVaR decelerates with $U_{\beta}^{\phi}(X)$ increases. If $S$ is linear, the degree of conservation of SlideVaR changes with a constant speed.
\subsection{Complete structure of SlideVaR}
Based on the above previous work, we give the complete structure of SlideVaR.
\begin{definition}
Given two confidence levels $\alpha$ and $\beta$, $\alpha \geqslant \beta$, for loss $X \in \mathcal{X}$,  $SlideVaR_{\alpha,\beta}^{\phi}(X)$ is defined by:
\begin{eqnarray}
SlideVaR_{\alpha,\beta}^{\phi}(X) = S(U_{\beta}^{\phi}(X))\cdot CVaR_{\alpha}(X) + [1-S(U_{\beta}^{\phi}(X))]\cdot VaR_{\beta}(X).
\end{eqnarray}
$U_{\beta}^{\phi}(X)$ is called tail thickness which is defined by:
\begin{eqnarray}
U_{\beta}^{\phi}(X):=\int_0^1F^{-1}_X(p)\phi_{\beta}(p)dp.
\end{eqnarray}
The risk aversion function $\phi_{\beta} \in L^1([0,1])$ is
\begin{eqnarray}
\phi_{\beta}(p) = \left\{
\begin{aligned}
& 0  					&\,\, \text{if }\,\, & p < \beta\\
& \phi(p) 				&\,\, \text{if }\,\, & p \geqslant \beta
\end{aligned} \right. ,
\end{eqnarray}
where $\phi \in L^1([F_{X}^{-1}(\beta),1])$ satisfies: (1) $\phi$ is positive, (2) $\phi$ is non-decreasing, (3) $|| \phi ||=1$. 
$S(x)$ is a normalization function which is defined by:
\begin{eqnarray}
S(x): [min(X),max(X)] \to [0,1].
\end{eqnarray}
\end{definition}\par

\subsection{The properties of SlideVaR}	
	As for mathematical properties, different researchers hold different opinions and desirable properties differ with the intended use for a risk measure. Drapeau\cite{Drapeau2013} emphasizes that a risk measure should satisfy the following two properties at the first: diversification and monotonicity. Diversification informs people don't put all your eggs in one basket. Monotonicity means that if a loss is greater in any situation as another one then it should be more risky. Next, we will show that monotonicity is a global property of SlideVaR.
\begin{theorem}\label{mon_S}
Given the confidence levels $\alpha,\beta$ and risk aversion function $\phi_{\beta}$, $\forall X_1 \geqslant X_2 \in \mathcal{X}$, 
\begin{eqnarray}
SlideVaR_{\alpha,\beta}^{\phi}(X_1) 	\geqslant SlideVaR_{\alpha,\beta}^{\phi}(X_2).
\end{eqnarray}
\end{theorem}\par			
	Actually, compared with monotonicity, the definition of diversification is controversial. For instance,  Artzner et al.\cite{Artzner1999} emphasize the sub-additivity but Goovaerts et al.\cite{darkiewicz2003} and Dhaene et al.\cite{dhaene2008} argue that this property may lead to undesirable situations. Drapeau \cite{Drapeau2013} embodies diversification as quasiconvexity. In this article, we still study sub-additivity. 
\begin{theorem}\label{sub_1}Given the normalization $S(x)$, let 
\begin{eqnarray}
\mathcal{D} = \{ x | S(x)=1\}
\end{eqnarray}
and
\begin{eqnarray}
\tilde{\mathcal{X}}= \{ X | U_{\beta}^{\phi}(X) \in \mathcal{D},X \in \mathcal{X} \}.
\end{eqnarray}
$\forall X,Y \in \tilde{\mathcal{X}}$, 
\begin{eqnarray}
SlideVaR_{\alpha,\beta}^{\phi}(X+Y) \leqslant SlideVaR_{\alpha,\beta}^{\phi}(X) + SlideVaR_{\alpha,\beta}^{\phi}(Y). 
\end{eqnarray}
\end{theorem}
Theorem \ref{sub_1} has another expression which we call risk-tail sub-additivity.
\begin{theorem}\label{sub_2}[risk-tail sub-additivity]Given the normalization $S(x)$, let 
\begin{eqnarray}
\mathcal{D} = \{ x | S(x)=1\}
\end{eqnarray}
and
\begin{eqnarray}
\tilde{\mathcal{X}}= \{ X | U_{\beta}^{\phi}(X) \in \mathcal{D},X \in \mathcal{X} \}.
\end{eqnarray}
Let
\begin{eqnarray}
\mathcal{X}^{\ast}= \{ X | U_{\beta}^{\phi}(X) = \min \limits_{X \in \tilde{\mathcal{X}}} U_{\beta}^{\phi}(X)\}.
\end{eqnarray}
$\forall X,Y\in \mathcal{X}$, if ${\exists}~X^{\ast} \in \mathcal{X}^{\ast}$~ s.t.~ $X,Y \geqslant X^{\ast}$, then
\begin{eqnarray}
SlideVaR_{\alpha,\beta}^{\phi}(X+Y) \leqslant SlideVaR_{\alpha,\beta}^{\phi}(X) + SlideVaR_{\alpha,\beta}^{\phi}(Y). 
\end{eqnarray}
\end{theorem}\par
Theorem \ref{sub_1} and Theorem \ref{sub_2} indicate that sub-additivity is a local property of SlideVaR. Considering the local region $\tilde{\mathcal{X}}$, 
\begin{theorem}
$\forall X \in \tilde{\mathcal{X}}$, if ${\exists}~Y$  s.t. $Y \geqslant X$, then $Y\in \tilde{\mathcal{X}}$.
\end{theorem}\par
It is the simple corollary of Theorem \ref{mon_U}. This theorem means that each element of $\tilde{\mathcal{X}}$ is no less risky than those outside the region and thus we call $\tilde{\mathcal{X}}$ risk-tail region.  Further, according to Theorem \ref{sub_2}, there exist a lower bound $\mathcal{X}^{\ast}$ of $\tilde{\mathcal{X}}$. For any losses which ``exceed'' the risk bound, SlideVaR satisfies sub-additivity. In other words, although sub-additivity is not a global property for SlideVaR, it still meets diversity requirements for high-risk assets or high-risk scenarios.\par
	Translation invariance give a monetary meaning to the risk measure which means that the risk of a position should be reduced by the amount of cash added to it. Positive homogeneity means that the risk increases linearly with the scale of the position. Nevertheless, F\"{o}llmer and Shied\cite{follmer2002} suggest that market risk may increase non-linearly with the value of the position. For example, it is possible that a liquidity risk be created when a position is multiplied by a sufficiently large factor\cite{Balbas2009}. In this terms, we have
\begin{theorem}\label{tran_S}
$\forall X \in \mathcal{X}$, if $a \in \mathcal{R}$ and $a \geqslant 0$, then
\begin{eqnarray}
SlideVaR_{\alpha,\beta}^{\phi}(X+a) 	\geqslant SlideVaR_{\alpha,\beta}^{\phi}(X)+a.
\end{eqnarray}
If $a \leqslant 0$, then
\begin{eqnarray}
SlideVaR_{\alpha,\beta}^{\phi}(X+a) 	\leqslant SlideVaR_{\alpha,\beta}^{\phi}(X)+a.
\end{eqnarray}
\end{theorem}
\begin{theorem}\label{pos_S}
$\forall X \in \mathcal{X}$, $\lambda \in \mathcal{R}$, if $\lambda \geqslant 1$, then
\begin{eqnarray}
SlideVaR_{\alpha,\beta}^{\phi}(\lambda X) 	\geqslant \lambda SlideVaR_{\alpha,\beta}^{\phi}(X).
\end{eqnarray}
If $0 \leqslant \lambda \leqslant 1$, then
\begin{eqnarray}
SlideVaR_{\alpha,\beta}^{\phi}(\lambda X) 	\leqslant \lambda SlideVaR_{\alpha,\beta}^{\phi}(X).
\end{eqnarray}
\end{theorem}
These two theorems mean that SlideVaR increases non-linearly with the scale of the loss. Moreover, SlideVaR is a convex risk measure in risk-tail region $\tilde{\mathcal{X}}$.
\begin{theorem}\label{convex_S}
$\forall \lambda \in[0,1]$, use the same notations as Theorem \ref{sub_1} and Theorem \ref{sub_2}, if 
\begin{eqnarray}
\left\{
\begin{aligned}
&~ X,Y \in \tilde{\mathcal{X}},\\
&~~~or\\
&~ X,Y\in \mathcal{X}~ and~ {\exists}~X^{\ast} \in \mathcal{X}^{\ast}~ s.t.~ X,Y \geqslant X^{\ast},
\end{aligned} \right.
\end{eqnarray}
we have
\begin{eqnarray}
SlideVaR_{\alpha,\beta}^{\phi}(\lambda X+(1-\lambda) Y) 	\leqslant \lambda SlideVaR_{\alpha,\beta}^{\phi}(X)+(1-\lambda) SlideVaR_{\alpha,\beta}^{\phi}(Y).
\end{eqnarray}
\end{theorem}\par
\section{Illustration}
	In order to demonstrate the proposed model, we provide simulate and empirical computation of VaR, CVaR, SlideVaR and GlueVaR. First of all, some basic settings are decided.\par
(a) Dominating confidence level $\alpha=0.99$, additional confidence level $\beta=0.95$. \par
(b) Risk aversion function $\phi(p)$ is an exponential function
\begin{eqnarray}
\phi_{\beta}(p) = \left\{
\begin{aligned}
& 0  					&\,\, \text{if }\,\, & p < \beta\\
& \dfrac{e^{\frac{p-1}{\gamma}}}{\gamma(1- e^{\frac{\beta-1}{\gamma}})}	&\,\, \text{if }\,\, & p \geqslant \beta
\end{aligned} \right.
\end{eqnarray}	
where $\gamma =0.2$.\par
(c) Using the same parameters as Ref\cite{Belles-Sampera2014a}: $\omega_1=\omega_2=\omega_3=\frac{1}{3}$. That is,
\begin{eqnarray}
GlueVaR_{0.95,0.99}(X) =\frac{1}{3}\cdot VaR_{0.95}(X)+ \frac{1}{3}\cdot CVaR_{0.95}(X)+\frac{1}{3}\cdot CVaR_{0.99}(X).
\end{eqnarray}

\subsection{Simulate computation of SlideVaR}
	Firstly, we set the normalization function $S(x)$ a linear function.
\begin{eqnarray}
S(x)=\left\{
\begin{aligned}
& 0        &\,\, \text{if }\,\, & x < a,\\
& \frac{1}{b-a}(x-a)        &\,\, \text{if }\,\, & a \leqslant x < b,\\
& 1        &\,\, \text{if }\,\, & b \leqslant x,	
\end{aligned} \right.
\end{eqnarray}
where $a = 20,~b=40$. Simulate data is generated randomly from a Gaussian mixture model which is 
\begin{eqnarray}
p(\theta)=\frac{1}{3}\mathcal{N}_1(\mu_1,\sigma^2_1)+\frac{2}{3}\mathcal{N}_2(\mu_2,\sigma^2_2).
\end{eqnarray}
We change the shape of the data distribution by altering the parameter $\mu_1,\sigma^2_1$ to observe the performance of SlideVaR.
\begin{figure}[htbp]
	\centering
	\includegraphics[width=0.9\linewidth]{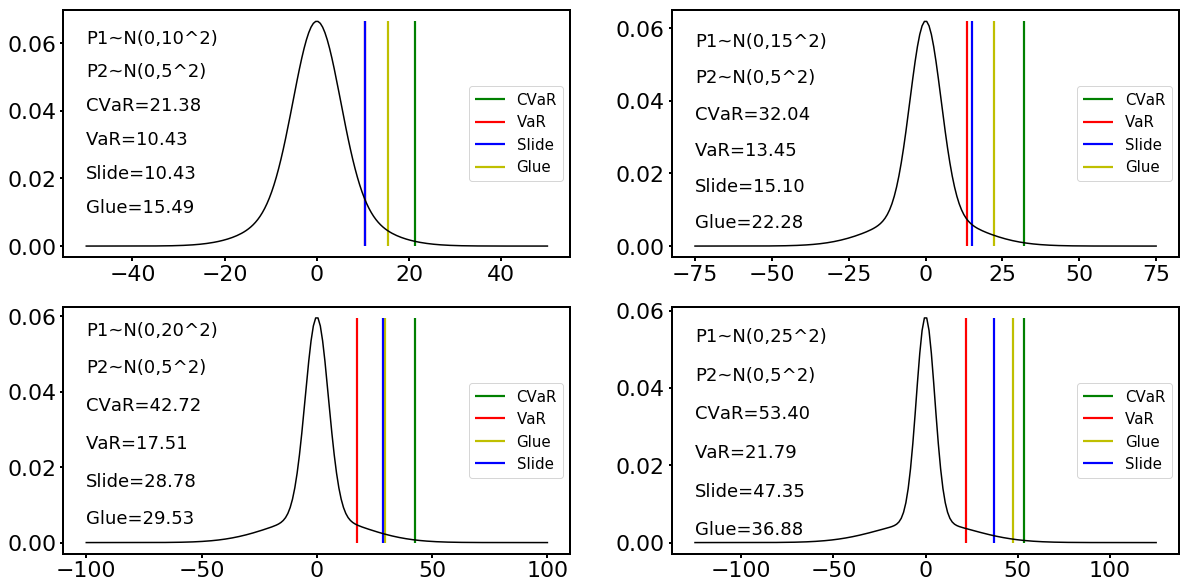} 
	\caption{Fix $\mu_1$ and change $\sigma_1$}  
	\label{fig1} 
\end{figure}\par
\begin{figure}[htbp]
	\centering
	\includegraphics[width=0.9\linewidth]{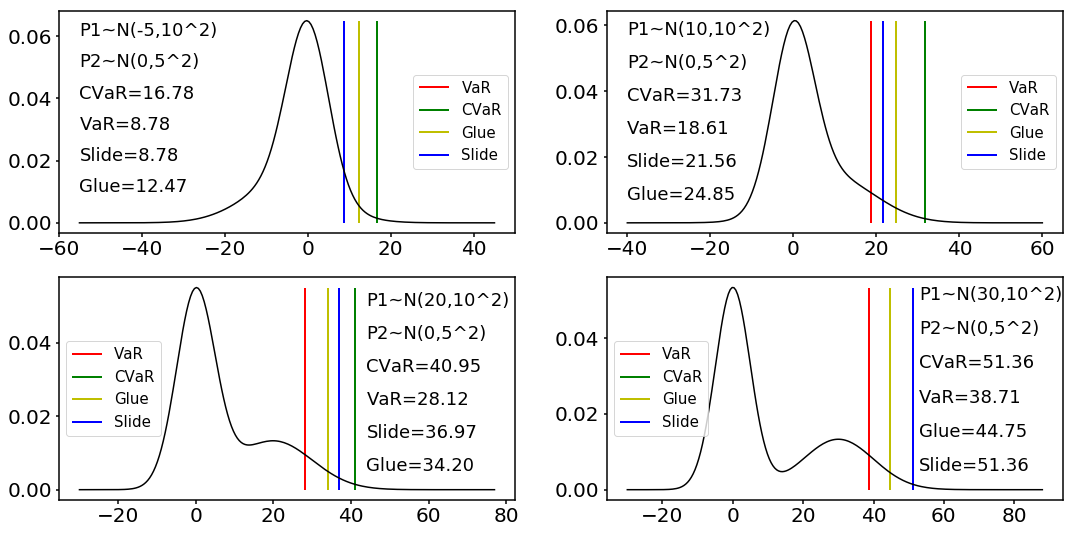} 
	\caption{Fix $\sigma_1$ and change $\mu_1$}  
	\label{fig2} 
\end{figure}\par
	First, we fix $\mu_1=0,\mu_2=0,\sigma_2=5$ and change $\sigma_1$ to 10, 15, 20, 25 as shown in Figure \ref{fig1}. Figure \ref{fig1} shows that as $\sigma_1$ increases, the peak of the distribution gradually becomes sharper, and accordingly, the tail of the distribution gradually becomes thicker. When $\sigma_1=10$ and $15$, SlideVaR is close to or even coincides with VaR and is less than GlueVaR. When $\sigma_1=20$, SlideVaR is between VaR and CVaR and is close to GlueVaR. When $\sigma_1=25$, SlideVaR is close to CVaR and greater than GlueVaR. Therefore, SlideVaR is becoming more and more conservative with $\sigma_1$ increases.\par
	Second, we fix $\sigma_1=10,\mu_2=0,\sigma_2=5$ and change $\mu_1$ to -5, 10, 20, 30 as shown in Figure \ref{fig2}. Figure \ref{fig2} shows that as $\mu_1$ increases, the tail of the distribution gradually becomes thicker and even shows bimodal forms(the last two pictures). When $\mu_1=-5$, SlideVaR is close to VaR and is less than GlueVaR. When $\mu_1=10$ and $20$, SlideVaR is between VaR and CVaR and is close to GlueVaR. When $\mu_1=30$, SlideVaR coincides with CVaR and greater than GlueVaR. Therefore, SlideVaR is becoming more and more conservative with $\mu_1$ increases.\par
\subsection{Empirical computation of SlideVaR}
The normalization function $S(x)$ for empirical data is
\begin{eqnarray}
S(x)=\left\{
\begin{aligned}
& 0        &\,\, \text{if }\,\, & x < a,\\
& \frac{1}{b-a}(x-a)        &\,\, \text{if }\,\, & a \leqslant x < b,\\
& 1        &\,\, \text{if }\,\, & b \leqslant x,	
\end{aligned} \right.
\end{eqnarray}
where $a = 1,~b=4$. We consider two financial markets i.e. Chinese(000001.SH from 1999 to 2018) and American(SPX.GI from 1999 to 2018)  securities market. We use historical simulation to estimate VaR, CVaR, SlideVaR and GlueVaR. The width of windows is 250 observations.\par
\begin{figure}[htbp]
	\centering
	\includegraphics[width=0.9\linewidth]{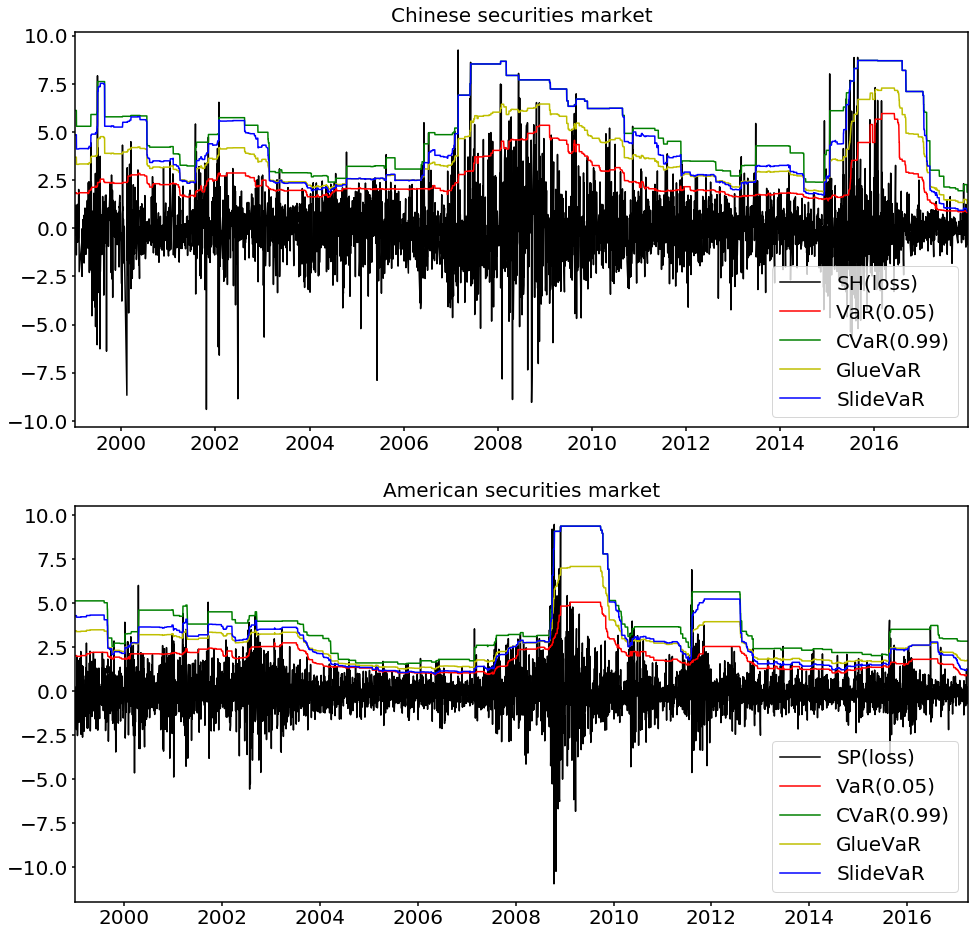} 
	\caption{Two financial markets}  
	\label{fig3} 
\end{figure}\par
	Figure \ref{fig3} shows that, in 1999, 2002, the periods from 2007 to 2009 and from 2015 to 2016, Chinese market suffered significant crisis. In these periods VaR underestimated extreme losses obviously and CVaR covered most losses. SlideVaR was very close to or even coincided with CVaR. Because GlueVaR was in the middle of VaR and CVaR, GlueVaR was still not as good as SlideVaR, although it performed better than VaR. A similar situation also occurred in American market in 2009 and 2011. In these periods American market also suffered significant crisis and SlideVaR covered more extreme losses than VaR and GlueVaR. Therefore, when a crisis strikes, SlideVaR has sufficient capacity to cover risks. This puts the investors who use SlideVaR to manage risk to suffer even smaller losses than ones who use VaR or GlueVaR.\par
	There were some differences in 2000, 2002 and the periods from 2015 to 2016. In these periods, the crisis was not particularly serious. SlideVaR was in the middle of VaR and CVaR, therefore SlideVaR and GlueVaR had similar performance. From 2003 to 2007 and from 2012 to 2015, SlideVaR's performance in the Chinese market was also similar to that of GlueVaR. Although Chinese market did not suffer any crisis during this period, SlideVaR was still significantly higher than VaR due to the high volatility of market. Therefore, SlideVaR and GlueVaR have similar performance for situations where the crisis is not very serious, or where the market is generally relatively high.\par
	From 2004 to 2007 and from 2013 to 2015, because of the low volatility of American market, SlideVaR was lower than GlueVaR and very close to VaR. This means that SlideVaR was more economical than GlueVaR and CVaR. During this period, investors using SlideVaR would face flexible risk control standard and less risk reserve capital allocation. This gave them more freedom of operation and more adequate capital to improve profitability. In other periods of the two markets, the market transitioned between crisis and stable periods. If market gets better, SlideVaR will cross over GlueVaR from below. If market gets worse, SlideVaR will cross under GlueVaR from above.\par
	In general, when a crisis strikes, SlideVaR is as safe as CVaR. SlideVaR is as economical as VaR when the market returns to a stable state.  When the market is in the middle, SlideVaR and GlueVaR behave similarly. When the market is in a transitional period, SlideVaR will follow the market to adjust its risk attitude adaptively. Therefore, SlideVaR has obvious advantages in markets where the state changes frequently.\par

\section{Conclusion}
	To find a trade-off between profitability and prudence, financial practitioners need to choose appropriate risk measures. VaR and CVaR are two of the most widely used methods. Simplicity is an important advantage of VaR. However, VaR may underestimate some certain catastrophic losses seriously and dose not satisfy the sub-additivity. CVaR was proposed as an improved method of which properties are more attractive than those of VaR: sub-additivity and convexity. Nevertheless, CVaR is difficult to calculate and more conservative and expensive than VaR. More importantly, CVaR still ignores the human's subjective risk attitude. Two key points are: Firstly, investors' risk attitudes under uncertainty conditions should be an important reference for risk measures. Secondly, the risk attitudes are not absolute. For different market performance, investors have different risk attitudes. For the first key point, Wang\cite{Wang1995,Wang1997} proposed distortion risk measure and Acerbi\cite{acerbi2001} proposed spectral risk measure. The distortion function $g$ and the risk aversion function $\phi$ appears as the instrument by which an investor can express his subjective attitude. GlueVaR was proposed by Belles-Sampera\cite{Belles-Sampera2014a} to devise a risk measure that lies somewhere between VaR and CVaR. It is a special distortion risk measure. However, for the second key point, these models ignore the impact of the market environment because their risk attitudes of are static. Frittelli\cite{Frittelli2014} proposed $\Lambda$VaR. In terms of risk attitudes, the tolerance level of $\Lambda$VaR will be modified as the loss distribution changes. However, $\Lambda$VaR can't describe how does the market environment affect risk attitudes. Besides, $\Lambda$VaR and VaR may lead to the same distasteful outcomes in some extreme situations.\par
	We proposed a new risk measure named SlideVaR. SlideVaR can be expressed as a combination of VaR and CVaR with two confidence levels. The weights of components are determined by $S(U_{\alpha}^{\phi}(X))$ which we proposed to measure the thickness of the tail of the distribution. $\phi$ reflects how investors assess the market situation and $S$ describe how investors' assessment influence their risk attitudes. Therefore, SlideVaR sufficiently considers different subjective attitudes of investors and sufficiently reflects the impact of market changes on investor attitudes. Moreover, SlideVaR has a relatively simple and intuitive form of expression for practical application.\par
	SlideVaR satisfies some of the good mathematical properties. First, monotonicity is a global property of SlideVaR. Next, we proposed the concept of risk-tail region i.e. each element of this region is no less risky than those outside the region. Accordingly, we proposed risk-tail sub-additivity and prove that SlideVaR satisfies it. Then, SlideVaR increases non-linearly with the loss increases. Finally, we illustrate that SlideVaR is convex in the risk-tail region. That is to say that, although sub-additivity and convexity are not global properties, SlideVaR still meets diversity requirements for high-risk assets or high-risk scenarios.\par
	In the illustration Section, we show that, SlideVaR is as safe as CVaR when a crisis strikes. When the market returns to a stable state, SlideVaR is as economical as VaR. SlideVaR and GlueVaR behave similarly when the market is in the middle. When the market is in a transitional period, SlideVaR will follow the market to adjust its risk attitude adaptively. In conclusion, SlideVaR has obvious advantages in markets where the state changes frequently.\par

\section{Appendix}
\subsection{Proof of theorem \ref{mon_S}}
\begin{proof}
According to Theorem \ref{mon_U}, for $X_1 \geqslant X_2$, we have
\begin{eqnarray}
S(U_{\beta}^{\phi}(X_1)) \geqslant  S(U_{\beta}^{\phi}(X_2)).
\end{eqnarray}
Because CVaR and VaR satisfy monotonicity, 
\begin{eqnarray}
CVaR_{\alpha}(X_1) \geqslant CVaR_{\alpha}(X_2),VaR_{\beta}(X_1) \geqslant VaR_{\beta}(X_2).
\end{eqnarray}
Therefore,
\begin{eqnarray}
\begin{aligned}
& SlideVaR_{\alpha,\beta}^{\phi}(X_1) - SlideVaR_{\alpha,\beta}^{\phi}(X_2) \\
=& S(U_{\beta}^{\phi}(X_1))\cdot CVaR_{\alpha}(X_1) - S(U_{\beta}^{\phi}(X_2))\cdot CVaR_{\alpha}(X_2)  \\
&~~~~+ [1-S(U_{\beta}^{\phi}(X_1))]\cdot VaR_{\beta}(X_1)- [1-S(U_{\beta}^{\phi}(X_2))]\cdot VaR_{\beta}(X_2) \\
\geqslant& S(U_{\beta}^{\phi}(X_2))\cdot [CVaR_{\alpha}(X_1) - CVaR_{\alpha}(X_2)] \\
&~~~~+ [1-S(U_{\beta}^{\phi}(X_1))]\cdot [VaR_{\beta}(X_1) - VaR_{\beta}(X_2)] \geqslant 0.
\end{aligned}
\end{eqnarray}
\end{proof}	 
\subsection{Proof of theorem \ref{sub_1}}
\begin{proof}
Because
\begin{eqnarray}
\forall X,Y \in \tilde{\mathcal{X}}, S(U_{\beta}^{\phi}(X))=S(U_{\beta}^{\phi}(Y))=1,
\end{eqnarray}
then
\begin{eqnarray}
SlideVaR_{\alpha,\beta}^{\phi}(X) = CVaR_{\alpha}(X),SlideVaR_{\alpha,\beta}^{\phi}(Y) = CVaR_{\alpha}(Y).
\end{eqnarray}
Because CVaR satisfies sub-additivity,
\begin{eqnarray}
CVaR_{\alpha}(X+Y) \leqslant CVaR_{\alpha}(X)+CVaR_{\alpha}(Y).
\end{eqnarray}
According to (\ref{mon_2}), we have
\begin{eqnarray}
\begin{aligned}
& SlideVaR_{\alpha,\beta}^{\phi}(X+Y) \leqslant CVaR_{\alpha}(X+Y) \\
&~~~~\leqslant CVaR_{\alpha}(X)+CVaR_{\alpha}(Y) = SlideVaR_{\alpha,\beta}^{\phi}(X)+SlideVaR_{\alpha,\beta}^{\phi}(Y).
\end{aligned}
\end{eqnarray}
\end{proof}
\subsection{Proof of theorem \ref{sub_2}}
\begin{proof}
It is the simple corollary of Theorem \ref{mon_U} and Theorem \ref{sub_1}.
\end{proof}
\subsection{Proof of theorem \ref{tran_S}}	
\begin{proof}
Because $U_{\beta}^{\phi}(X)$ is a special spectral risk measure, it satisfies translation invariance. i.e. $\forall a \in \mathcal{R}$,
\begin{eqnarray}
U_{\beta}^{\phi}(X+a) = U_{\beta}^{\phi}(X)+a.
\end{eqnarray}
Then,
\begin{eqnarray}
\begin{aligned}
& SlideVaR_{\alpha,\beta}^{\phi}(X+a)- SlideVaR_{\alpha,\beta}^{\phi}(X) - a\\
=& S(U_{\beta}^{\phi}(X)+a)\cdot CVaR_{\alpha}(X) + [1-S(U_{\beta}^{\phi}(X)+a)]\cdot VaR_{\beta}(X) \\
&~~~~- S(U_{\beta}^{\phi}(X))\cdot CVaR_{\alpha}(X) + [1-S(U_{\beta}^{\phi}(X))]\cdot VaR_{\beta}(X)\\
=& [S(U_{\beta}^{\phi}(X)+a) - S(U_{\beta}^{\phi}(X))] \cdot [ CVaR_{\alpha}(X) - VaR_{\beta}(X)].
\end{aligned}
\end{eqnarray}
If $a \geqslant 0$, then $S(U_{\beta}^{\phi}(X)+a) \geqslant S(U_{\beta}^{\phi}(X))$ and then 
\begin{eqnarray}
SlideVaR_{\alpha,\beta}^{\phi}(X+a) \geqslant  SlideVaR_{\alpha,\beta}^{\phi}(X) + a.
\end{eqnarray}
If $a \leqslant 0$, then $S(U_{\beta}^{\phi}(X)+a) \leqslant S(U_{\beta}^{\phi}(X))$ and then 
\begin{eqnarray}
SlideVaR_{\alpha,\beta}^{\phi}(X+a) \leqslant  SlideVaR_{\alpha,\beta}^{\phi}(X) + a.
\end{eqnarray}
\end{proof}	
\subsection{Proof of theorem \ref{pos_S}}
\begin{proof}
Because $U_{\beta}^{\phi}(X)$ is a special spectral risk measure, it satisfies positive homogeneity. i.e.  $\forall \lambda \geqslant 0$,
\begin{eqnarray}
U_{\beta}^{\phi}(\lambda X) = \lambda U_{\beta}^{\phi}(X).
\end{eqnarray}
Then,
\begin{eqnarray}
\begin{aligned}
& SlideVaR_{\alpha,\beta}^{\phi}(\lambda X)- \lambda SlideVaR_{\alpha,\beta}^{\phi}(X) \\
=&  \lambda \cdot \{ S(U_{\beta}^{\phi}(\lambda X))\cdot CVaR_{\alpha}(X) + [1-S(\lambda U_{\beta}^{\phi}(X))]\cdot VaR_{\beta}(X) \\
&~~~~- S(U_{\beta}^{\phi}(X))\cdot CVaR_{\alpha}(X) + [1-S(U_{\beta}^{\phi}(X))]\cdot VaR_{\beta}(X) \} \\
=& \lambda \cdot [S(\lambda U_{\beta}^{\phi}(X)) - S(U_{\beta}^{\phi}(X))] \cdot [ CVaR_{\alpha}(X) - VaR_{\beta}(X)].
\end{aligned}
\end{eqnarray}
If $\lambda \geqslant 1 $, then $S(\lambda U_{\beta}^{\phi}(X)) \geqslant S(U_{\beta}^{\phi}(X))$ and then 
\begin{eqnarray}
SlideVaR_{\alpha,\beta}^{\phi}(\lambda X) \geqslant  \lambda SlideVaR_{\alpha,\beta}^{\phi}(X).
\end{eqnarray}
If $0 \leqslant \lambda \leqslant 1$, then $S(\lambda U_{\beta}^{\phi}(X)) \leqslant S(U_{\beta}^{\phi}(X))$ and then 
\begin{eqnarray}
SlideVaR_{\alpha,\beta}^{\phi}(\lambda X) \leqslant  \lambda SlideVaR_{\alpha,\beta}^{\phi}(X) .
\end{eqnarray}
\end{proof}
\subsection{Proof of theorem \ref{convex_S}}
\begin{proof}
It is the simple corollary of Theorem \ref{sub_1}, Theorem \ref{sub_2} and Theorem \ref{non-linear_1}.
\end{proof}
	
\bibliographystyle{unsrt}
\bibliography{ref}

\end{document}